\documentclass[10pt, final, conference]{IEEEtran}
\IEEEoverridecommandlockouts
\usepackage{cite}
\usepackage{mathtools,amssymb,amsfonts}
\usepackage{siunitx}
\usepackage{algorithmic}
\usepackage{graphicx}
\usepackage{textcomp}
\usepackage{xcolor}
\usepackage{tabto}
\usepackage{pgfplots}					
\usepackage{pgfplotstable}
\usepgfplotslibrary{statistics}		
\pgfplotsset{width=7cm,compat=newest}	
\usetikzlibrary{shapes.geometric}
\newcommand\TP{\mathit{TP}}
\newcommand\FN{\mathit{FN}}
\newcommand\FP{\mathit{FP}}

\usepackage{todonotes}

\usetikzlibrary{external}
\def\BibTeX{{\rm B\kern-.05em{\sc i\kern-.025em b}\kern-.08em
    T\kern-.1667em\lower.7ex\hbox{E}\kern-.125emX}}
\begin{document}

\title{
Automated Ground Truth Estimation of Vulnerable Road Users in Automotive Radar Data Using GNSS
}

\author{\IEEEauthorblockN{Nicolas Scheiner, Nils Appenrodt, J\"urgen Dickmann}
\IEEEauthorblockA{\textit{Environment Perception} \\
\textit{Daimler AG}\\
Ulm, Germany \\
\{nicolas.scheiner, nils.appenrodt, juergen.dickmann\}@daimler.com}
\and
\IEEEauthorblockN{Bernhard Sick}
\IEEEauthorblockA{\textit{Intelligent Embedded Systems} \\
\textit{University of Kassel}\\
Kassel, Germany \\
bsick@uni-kassel.de}
}

\maketitle
\thispagestyle{plain}  
\pagestyle{plain}  

\begin{abstract}
Annotating automotive radar data is a difficult task.
This article presents an automated way of acquiring data labels which uses a highly accurate and portable global navigation satellite system (GNSS).
The proposed system is discussed besides a revision of other label acquisitions techniques and a problem description of manual data annotation.
The article concludes with a systematic comparison of conventional hand labeling and automatic data acquisition.
The results show clear advantages of the proposed method without a relevant loss in labeling accuracy.
Minor changes can be observed in the measured radar data, but the so introduced bias of the GNSS reference is clearly outweighed by the indisputable time savings.
Beside data annotation, the proposed system can also provide a ground truth for validating object tracking or other automated driving system applications.
\end{abstract}

\section{Introduction}
Autonomous transportation is a driving force in current automotive research.
Perception tasks for automotive applications were traditionally based on classical signal processing methods.
Nowadays, a clear shift towards techniques originating from the field of artificial intelligence can be observed.
The machine learning methods used for these tasks usually follow a supervised learning paradigm.
In supervised learning, a machine learning model is training on annotated data samples to adjust the model coefficients to be able to make good predictions for previously unseen data.
The amount of annotated data required for training depends on both the complexity of used models as well as the difficulty of the task itself.
On the way towards autonomous driving it is most likely that vast amounts of data are required to ensure excellent environmental perception under varying measurement conditions.
Data acquisition is always a tedious and expensive task, in particular for abstract data representations.
A human labeler is simply not as accustomed to radar measurements as to image data, for example.
One solution to this problem is to find models that require less data to converge, i.e., using classification models with small amount of training coefficients as in \cite{Scheiner2018}.
Moreover, techniques such as active learning were found to be able to reduce the required amount of data for radar-based classification tasks \cite{Winterling2017}.

A different way to address the increasing needs for data is the simplification of the labeling process.
A common approach is the utilization of cross-modalities between sensors for transforming data annotations in one domain to another, e.g., camera images to lidar point clouds as shown in \cite{Piewak2018}.
In this case, both sensors have a similar behavior with regards to their way of perceiving a scene, i.e., both detect electromagnetic waves in or near the visible frequency range.
Automotive radar sensors typically operate in the \SI{24}{GHz} or \SI{77}{GHz} bands and, thus, can propagate through many obstacles that would limit other sensors.
The enhanced view on partially occluded objects is very difficult to infer for a label transformation process.

\begin{figure}[t]
\centerline{\includegraphics[width=\columnwidth]{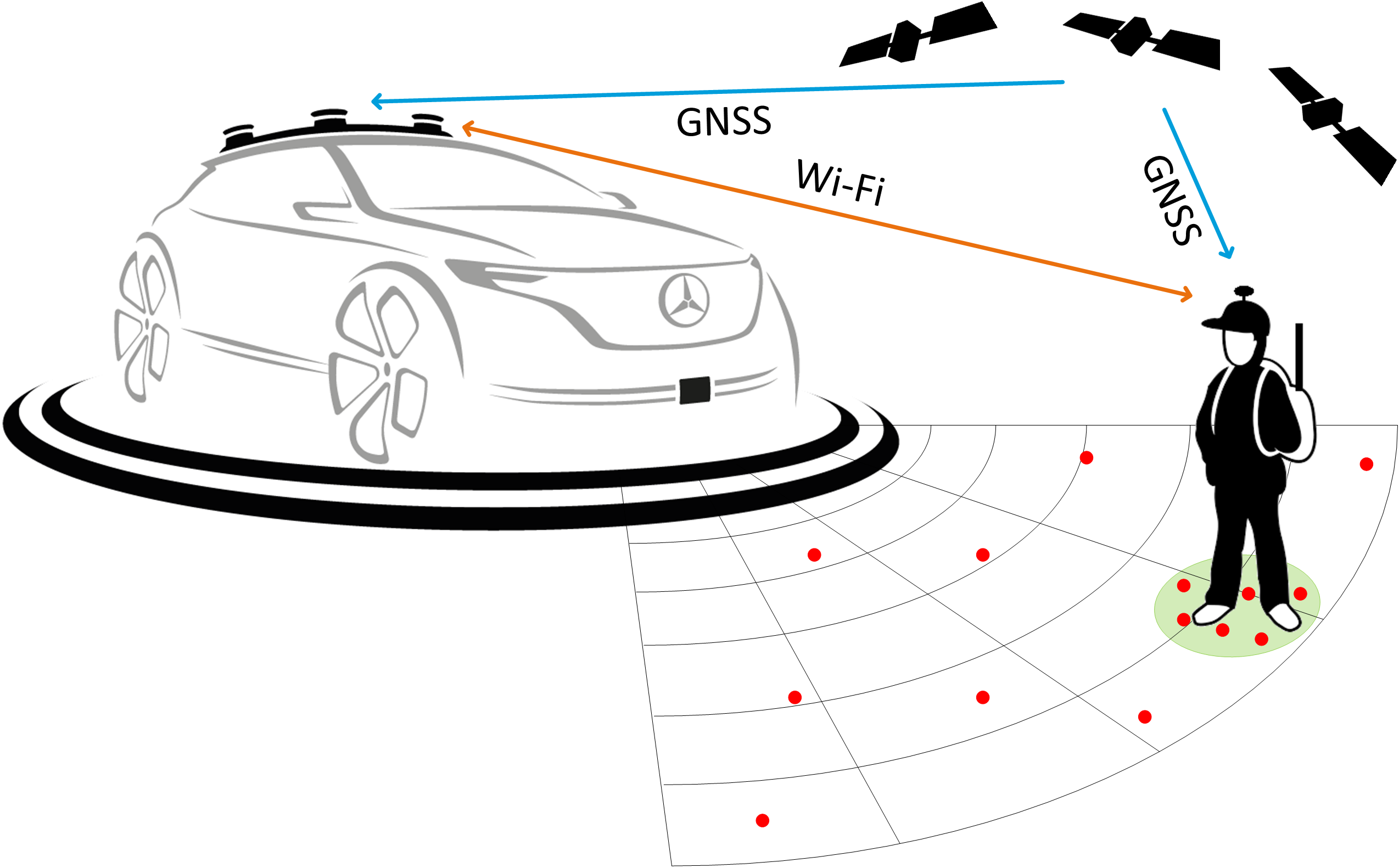}}
\caption{Automated label generation with a GNSS reference system: global positions of ego-vehicle and VRU are used to automatically assign labels to radar data in close proximity to the VRU's location.}
\label{fig:sys_overview}
\end{figure}
When allowing to limit parts of the data generation process to instructed scenarios the labeling process can also be facilitated with the use of reference sensors.
To this end, this article equips vulnerable road users (VRU) with hand-held global navigation satellite system (GNSS) modules that are referenced to another GNSS module mounted on a vehicle (cf. Fig. \ref{fig:sys_overview}).
\begin{figure*}[htb!]
\centerline{\includegraphics[width=\textwidth]{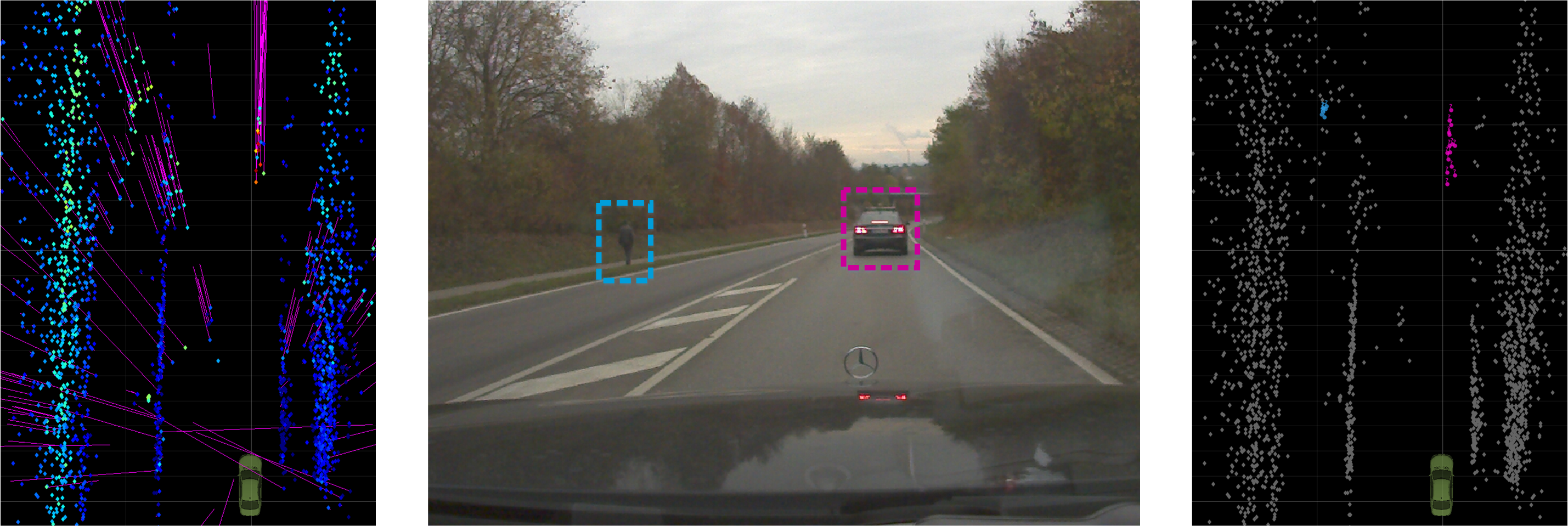}}
\caption{Left: labeling view on detection-level radar data. Amplitude values are color encoded, ego-motion compensated radial velocities are depicted by the magenta lines with their length representing the speed. The middle image is from a documentation camera. The two objects of interest are highlighted here for better visibility. On the right is the target view, where corresponding radar detections have been grouped and assigned with the correct label.}
\label{fig:radar_labeling}
\end{figure*}
This makes it possible to determine the relative positioning of vehicle and observed object.
An adaptive selection algorithm associates radar data points to an area around the estimated observed position in order to provide a ground truth for the measurement data.
The auto-labeled data is then compared to manually labeled measurements in both labeling quality and differences in measurement values due to the reference system.
Final results clearly state the validity of the method with only minor drawbacks when compared to manual labeling.
Although, the main focus of this article is automated data annotation, the additional information gained from a GNSS module is also very useful for other applications such as object tracking.

The article is structured as follows:
Sections \ref{sec:labeling} and \ref{sec:system} discuss the conventional labeling process and the proposed system.
In Section \ref{sec:exp}, the auto-labeling method is evaluated with regards to its accuracy and influence on the measurement quality based on a series of experiments.
Finally, Section \ref{sec:conclusion} concludes the topic and gives prospects for future work.

\section{Radar Labeling}\label{sec:labeling}
To motivate the topic and to provide a baseline for comparison, a short introduction to conventional radar labeling is given in this section.
As stated above, annotating automotive radar measurements is cumbersome.
Depending on the complexity of the scenario it may take several hours to label only a few seconds of radar data.
Manual annotation requires trained experts, as graphical data representations are nowhere as evident as, e.g. video data.
The expert can be presented with different representations of the data.
Fig. \ref{fig:radar_labeling} gives an example of a possible view on a snapshot of a typical radar scene along with the corresponding documentation camera image and the target view.
The data is presented to the labeler on detection level, i.e., prefiltered by a constant false alarm rate (CFAR) \cite{richards2005} detector.
The reflections are resolved in range, angle, and radial Doppler velocity.
The chosen representation for Fig. \ref{fig:radar_labeling} displays ego-motion compensated radial velocities of all data points by arrows pointing towards the sensor.
The reflection amplitudes are color encoded, however, these settings might be adjusted to meet the needs of the current scene.
By comparing the radar image with the camera view, objects are identified and individual points are assigned a label.
For the sake of better visibility only excerpts of the vehicle's field of view are used here, the whole scene will be even harder to match with the image.
Estimating the correct distance in the radar image can be eased by using a stereo camera or lidar sensor as reference.
In return, understanding the range image or lidar point cloud becomes less intuitive.
Thus, the advantage over a series of regular 2D color images which can be used for going back and forth in time is not essential.

\section{Proposed Auto-Labeling System}\label{sec:system}
In order to deal with the increasing demands for big data sets, this article suggests to gather big parts of the required training data by recording instructed traffic participants for which the exact location is known at all times.
Literature provides different kinds of localization systems. 
Only those are relevant which are portable, can easily be carried by a pedestrian, and allow to move out of sight of the vehicle without losing track to the location.
Interesting approaches combine different sensory data from smartphones as in \cite{Agrawal2013} or use ultra-wideband radio signals from fixed locations for position estimation \cite{Ledergerber2015}.
While the first approach with smartphones is very appealing due to its low hardware requirements, an accuracy of $2-$\SI{4}{m} is not precise enough for many radar scenes.
Instead, this article relies on a GNSS architecture which is somewhat similar to the method described in \cite{Ledergerber2015}.

\subsection*{Position And Motion Tracking Hardware}
\begin{figure}
\centerline{\includegraphics{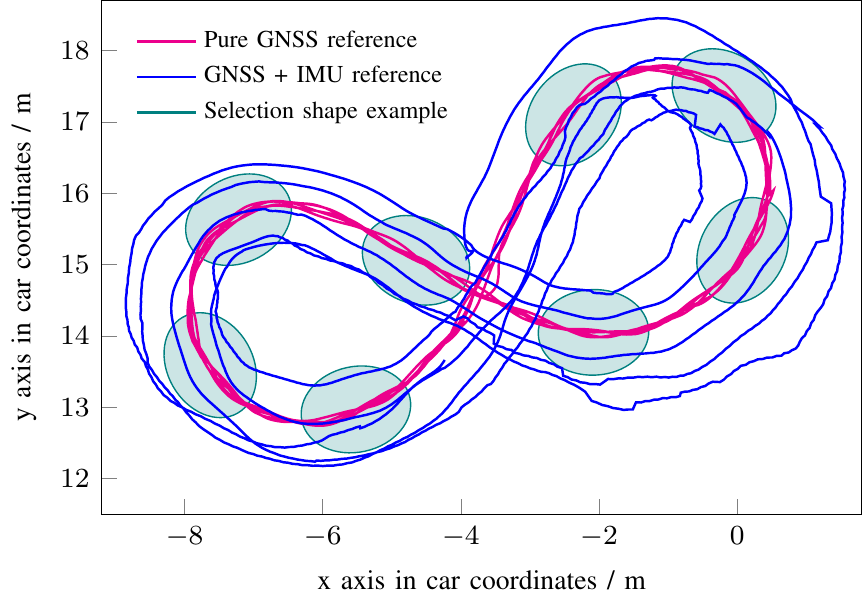}}
\caption{Five repetitions of estimated reference system trajectories based on GNSS position with and without usage of IMU smoothed with a moving average filter of length 9. Examples for an ellipse-based selection area are given for the more stable pure GNSS trajectory.}
\label{fig:eight}
\end{figure}

The proposed auto-labeling system consists of the following components:
vehicle and VRUs are each equipped with a device combining a GNSS receiver and an inertial measurement unit (IMU) for orientation estimation.
VRUs comprise pedestrians and cyclists for this article.
The communication between car and VRU is handled via Wi-Fi.
Instead of using Wi-Fi, the GNSS positions and times of both VRU and ego-vehicle could also be stored and processed offline as an alternative or additional approach.
In general, this could increase reliability for, e.g., fast drive-by maneuvers.
Due to the limited measurement distance of the radar sensors and for real-time surveillance purposes a Wi-Fi data transmission was chosen.
The GNSS receivers use GPS and GLONASS satellites and real-time kinematic (RTK) positioning to reach centimeter-level accuracy.
RTK is a more accurate version of GNSS processing which uses an additional base station with known location in close distance to the desired position of the so-called rover \cite{Thomas2010}.
It is based on the assumption that most errors measured by the rover are essentially the same at the base station and can, therefore, be eliminated by using a correction signal that is sent from base station to rover.
All system components for the VRU system except the antennas are installed in a backpack including a power supply.
The GNSS antenna is mounted on a hat to ensure best possible satellite reception, the Wi-Fi antenna is attached to the backpack.
Especially for the ego-vehicle, a complete pose estimation (position + orientation) is necessary for correct annotation of global GNSS positions and radar measurements in sensor coordinates.
Moreover, both vehicle and VRU can benefit from a position update via IMU if the GNSS signal is erroneous or simply lost for a short period.
Experiments, however, did quickly reveal that the internal Kalman filter \cite{Kalman1960}, which fuses both signals in the GNSS + IMU unit, is not well equipped for unsteady movements of VRUs, especially not for pedestrians.
Exemplary trajectories of combined GNSS + IMU positioning versus pure GNSS can be found in Fig. \ref{fig:eight}, along with some examples of the data selection area which will be explained in the remainder of this section.
In Fig. \ref{fig:eight} only the pure GNSS trajectory remains on the preset eight-shaped course while the combined GNSS + IMU trajectory quickly accumulates positioning errors and drifts away.
For this reason only the GNSS position is used for further experiments.
\begin{figure}[b]
\centering
\includegraphics[width=\columnwidth]{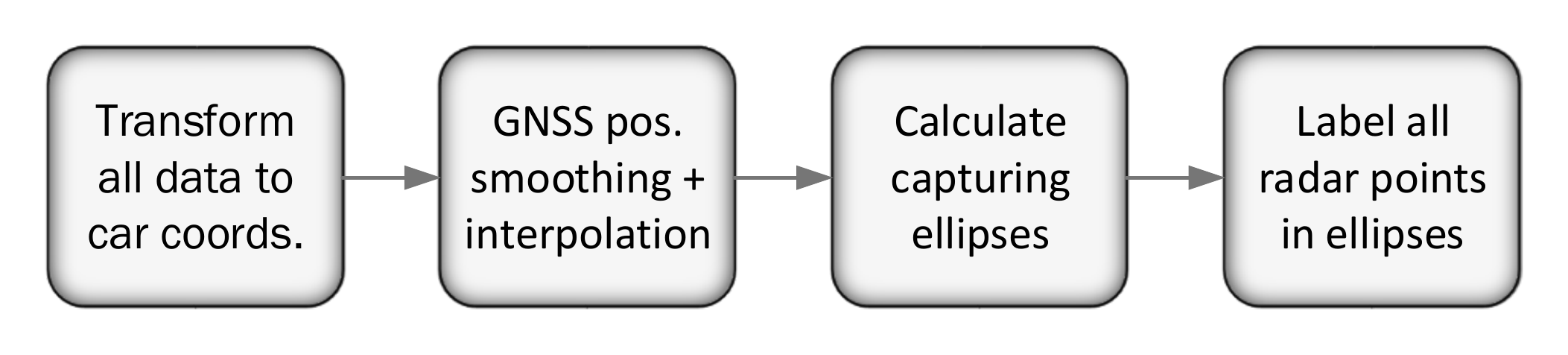}
\caption{Auto-labeling selection strategy processing steps}
\label{fig:proc_chain}
\end{figure}

\subsection*{Selection Strategy}
Once all data from GNSS and radar is captured, the particular VRU label has to be assigned to corresponding radar reflections.
A basic overview is given in Fig. \ref{fig:proc_chain}.
At first, all data needs to be transformed to a common coordinate system, e.g., car coordinates.
Then, GNSS information is smoothed with a moving average filter of length $9$ to remove jitter in the positions.
The length corresponds to roughly \SI{0.5}{sec} and is a trade-off between good smoothing characteristics and the expected interval of continuous VRU behavior.
For each timestamp of each radar measurement the GNSS position is estimated by cubic spline interpolation.
In the next step, an area around the position has to be defined. 
If ordinary cars were auto-labeled with this method a selection area could be easily defined as car orientation and outer dimensions can usually be estimated very precisely.
For VRUs, the selection area is more volatile, though.
A major component for the higher difficulty is the missing orientation due to inaccuracies in the IMU measurements as described above.
Also, swinging body parts or turning handlebars of cyclists, for example, complicate defining a fixed enclosing structure.
Hence, two different versions of surrounding shapes are proposed for the VRUs under consideration.
For a pedestrian, a Gaussian distribution is assumed, thus an ellipse is used with its major axis oriented in movement direction (yaw angle) $\phi$.
Major and minor axis of the ellipse are calculated from fixed minimum pedestrian dimensions (empirically determined: $\SI{1.5}{m} \times \SI{1.2}{m}$) plus a variable extra length for swinging body parts defined by its velocity $v$ and yaw rate $\dot{\phi}$:
\begin{equation}
\resizebox{.89\linewidth}{!}{%
  $\text{ax}_\text{maj.}=\begin{cases}
    \SI{1.5}{m} + \min(|v| \cdot \SI{}{sec}, \SI{1}{m}), & \text{if $v\geq0.05\frac{\SI{}{m}}{\SI{}{s}}$}.\\
    \SI{1.5}{m}, & \text{otherwise}.
  \end{cases}$%
  }
\end{equation}
\begin{equation}
\resizebox{.9\linewidth}{!}{%
  $\text{ax}_\text{min.}=\begin{cases}
    \SI{1.2}{m} + \min(|\dot{\phi}| \cdot 5\frac{\SI{}{m} \cdot \SI{}{sec}}{\SI{}{rad}}, \SI{1}{m}), & \text{if $v\geq0.05\frac{\SI{}{m}}{\SI{}{s}}$}.\\
    \SI{1.5}{m}, & \text{otherwise}.
  \end{cases}$%
  }
\end{equation}
Since the IMU of the reference system did not prove to be useful given its current configuration, $v$, $\phi$, and $\dot{\phi}$ are estimated based on regression lines calculated from consecutive GNSS position measurements centered around each sample with a maximum distance of \SI{0.25}{m}.
If the pedestrian pauses and the time difference to the next location in \SI{0.25}{m} distance exceeds a threshold level of \SI{2}{sec}, the velocity and yaw angle estimation are not stable anymore, hence a circle is used in this case.
The cyclist is labeled inside a rectangle with fixed length of \SI{2.5}{m} oriented in driving direction $\phi$ and width of \SI{1.2}{m} plus a variable amount based on its yaw rate.
\begin{equation}
     \text{width}_\text{rect} = \SI{1.2}{m} + \min(|\dot{\phi}| \cdot 5\tfrac{\SI{}{m} \cdot \SI{}{sec}}{\SI{}{rad}}, \SI{1}{m}).
\end{equation}
As bikes usually cannot turn without driving, the derivation of $\phi$ assumes constant continuation of the cyclist's orientation.
Therefore, no special treatment is required for stopping bikes.

Lastly, at each time step all radar detections that lie inside the defined regions are being assigned the corresponding label.

\section{System Evaluation}\label{sec:exp}

\subsection{Experiments}
A series of experiments were conducted to evaluate the performance of the proposed auto-labeling system.
In order to get measurements from all angles of the VRU, a track with the shape of an eight was marked on the ground for the test subjects to walk or drive on (cf. Fig \ref{fig:eight}).
Three different scenarios consisting of a total of over 9000 radar measurement cycles were evaluated:
\begin{enumerate}
\item Pedestrian walking at constant normal speed
\item Pedestrian walking around two bikes which are placed as obstacles in the bulges of the eight
\item Cyclist driving at approximately \SI{3}{\meter\per\second}
\end{enumerate}
The scenarios 1) and 3) were repeated five times with and five times without carrying the reference system.
Scenario 2) only serves as a reference for labeling accuracy, hence this experiment was only conducted with the GNSS backpack.
All measurements were hand-labeled by a human expert and additionally all measurements including a GNSS reference were labeled automatically with the proposed method.
The experiments were carried out by two chirp sequence radars operating at \SI{77}{GHz} with resolutions of roughly \SI{15}{\centi\meter} in range, \SI{2.4}{\degree} in azimuth angle, and \SI{0.17}{m/s} in radial velocity.

\subsection{Results}
Several indicators are important for comparing the proposed method with conventional manual labeling.
First, the accuracy of the method has to be compared against the ground truth obtained from manual labeling.
Second, the differences in measured values for VRU carrying or not carrying the reference system have to be estimated.

To determine the performance of the auto-labeling system, two scores were calculated.
Let $\TP$ (true positives) be the amount of correctly labeled points, $\FP$ (false positives) the incorrectly labeled points, and $\FN$ (false negatives) the amount of points that incorrectly have not been assigned a label.
Then, the precision of the method can be calculated as $\mathit{Pr} = \TP/(\TP+\FP) \in [0,1]$ and the recall is $\mathit{Re} = \TP/(\TP+\FN) \in [0,1]$.
The scores are calculated for all scenes including the GNSS reference.
The macro-averaged results, i.e., averaged individual scores yield a \textbf{precision} of \textbf{\SI{99.48}{\%}} and a \textbf{recall} of \textbf{\SI{99.66}{\%}}.
Please note, that it would certainly be possible to improve these scores for this data set by fine-tuning the parameters of the selection area.
This could, however, easily result in an overfitting on the given data set, i.e., a parameterization that would not generalize well on other data.
As the main goal of this test is the validation of the method itself, both values prove that the selection algorithm was parameterized reasonably.

\begin{figure}
\centerline{\includegraphics{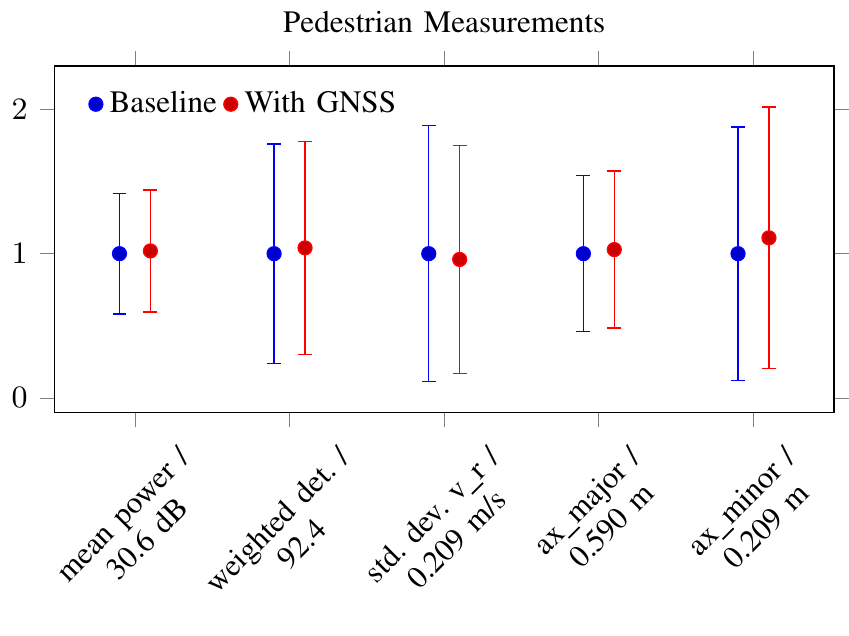}}
\centerline{\includegraphics{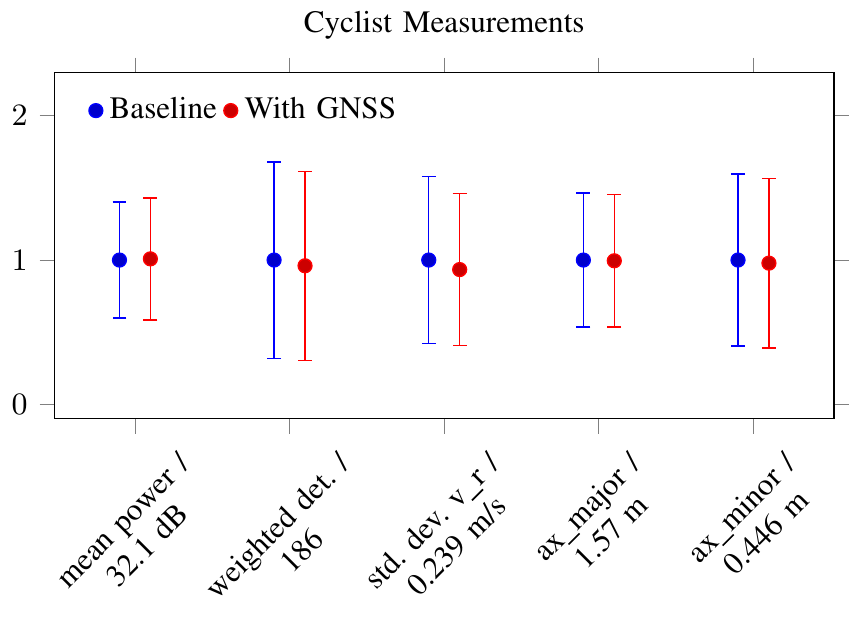}}
\caption{Experimental results for pedestrian (top) and cyclist (bottom) each with and without reference system. Measured values are compared in five categories. In each category all mean values and standard deviations are displayed and all values are normalized to the mean baseline value of the corresponding scenario.}
\label{fig:results}
\end{figure}

In order to determine how wearing the GNSS equipment alters measured values, manually labeled data of scenes 1) and 3) are compared for scenarios was and was not worn.
Important criteria for comparison are measured amplitudes, variations of Doppler values, the spatial extent, and the amount of detections per measurement.
Therefore, the mean reflected power, the standard deviation of Doppler values, the length of the major and minor axis of the \SI{95}{\%} confidence ellipse, and the amount of detections weighted by the mean distance to the sensor are calculated for each measurement cycle.
Scans from the two sensors are hereby treated as different measurements.
These features were previously found to be well-suited characteristics for the classification of VRUs in \cite{Scheiner2018}.
The reflected power is compensated for free-space path loss using $R^4$ correction where $R$ is the range between sensor and detection.
A reference target for estimating the radar cross-section is not available, therefore absolute power values have to be treated with care.
Fig. \ref{fig:results} displays the averaged results and their standard deviations.
For better displaying, all values are normalized to the mean values of the corresponding variable from the measurements without GNSS backpack.
Unpaired t-tests (significance level $\alpha = 0.05$) on the overall pedestrian data reveal a strong statistically significant difference in the length of the minor axis of the \SI{95}{\%} confidence ellipse (p-value $= 0.0006$).
No statistically significant differences can be found for the standard deviations in returned power, Doppler variation, weighted amount of detections or the confidence ellipse's major axis (p-values of $0.2075$, $0.1840$, $0.1373$, and $0.1458$).
For the bike, statistically significant differences can only be found for the standard deviation of Doppler values (p-value $= 0.0016$), whereas power, number of detections, as well as the major and minor axis are all not statistically significantly different (p-values of $0.5941$, $0.1160$, $0.7716$, and $0.3415$).
The increased significance of Doppler variations within one sensor scan when compared to the pedestrian is not an expected result.
One explanation for this behavior would be that the cyclist was simply moving differently during the measurements.
This hypothesis is backed by the not overly critical significance level and the difficulty in keeping the speed steady during long scenes.
In summary, the results are as expected.
All examined criteria except the variation in Doppler slightly increase when the reference system is worn by a pedestrian.
For a cyclist, measured values are even more similar in both cases.
Statistically significant differences can only be observed in the width of a pedestrian and the variations in Doppler of a cyclist.
The same effect on the elongation of a pedestrian should, however, also occur from any other ordinary backpack with e.g. a laptop inside.
Hence, the observed differences introduce a bias, but are unlikely to make them less relevant.

Despite good results, the proposed method also has some drawbacks which are more difficult to measure.
The necessity to wear the GNSS equipment during all measurements limits the number of VRUs to track at a time to the number of available backpacks.
This means that only instructed people can be auto-labeled and it is hard to make sure that they do not behave differently because they know that they are being monitored.
Also, data acquisition with multiple sensors might have problems, e.g. an image recognition algorithm might learn that pedestrians always have GNSS antenna on their head, which is obviously not true.
Keeping in mind the enormous time savings that can be made by using this method, the advantages clearly outweigh those flaws.
As an example, even the trivial scenes which were labeled for this article took an average of \SI{18}{min} for labeling per scene (real sequence time $\approx$\SI{2}{min} on average), even though just a single VRU with a very simple trajectory had to be labeled.
The difference becomes obvious when comparing the total of \SI{90}{min} of labeling effort with the approximated time of $5-$\SI{10}{min} to initialize an arbitrary number of GNSS modules which are ready for hours then.
For a final application it is proposed to use both auto-labeled and human-labeled data in order to get a huge database which is partially without the potential bias created by the reference system and the corresponding instructed VRUs.

\section{Conclusion}\label{sec:conclusion}
In this article, a method for automating radar data labeling of VRUs was proposed.
The system is based on the combination of two GNSS receivers mounted on the ego-vehicle and the VRU.
Radar data is automatically assigned a label if it falls within a close area around the VRU's GNSS location.
The selection area is determined by the kind of tracked VRU, i.e., pedestrian or cyclist, its speed and yaw rate.
Experiments prove the accuracy of the proposed method with precision and recall both over \SI{99}{\%}.
Though significant changes of measured values can be estimated for a pedestrian's elongation in width as well as the cyclist's variations of Doppler values, these changes are still not very big and should be easy to replicate with any ordinary backpack.
As every hour of automatic labeling may save days of labeling capacity, the trade-off seems negligible.
For the major part of all observed values, the differences are not statistically significant.
Nevertheless, a final recommendation is to use the proposed system as an addition to manual labeling for an efficient drastic enlargement of the database.
By also relying on conventionally annotated data it can then be assured to prevent from getting compromised by the small bias introduced by the proposed system.
In future work it is planned to use several GNSS backpacks for tracking multiple VRUs.
This involves adapting the selection strategy to cope with situations where selection areas overlap.
Furthermore, an improved fusion algorithm for IMU and GNSS shall be investigated to also benefit from the IMU data.
It is also planned to collect a lot more data with and separately also without GNSS reference.
By analyzing the performance of a classification algorithm with automatically generated training data and manually labeled test data the findings of this article shall be validated.

\section{Acknowledgment}
The research for this article has received funding from the European Union under the H2020 ECSEL Programme as part of the DENSE project, contract number 692449.

\bibliographystyle{IEEEtran}
\bibliography{IEEEabrv,mybibfile}

\begin{thebibliography}{1}
\providecommand{\url}[1]{#1}
\csname url@samestyle\endcsname
\providecommand{\newblock}{\relax}
\providecommand{\bibinfo}[2]{#2}
\providecommand{\BIBentrySTDinterwordspacing}{\spaceskip=0pt\relax}
\providecommand{\BIBentryALTinterwordstretchfactor}{4}
\providecommand{\BIBentryALTinterwordspacing}{\spaceskip=\fontdimen2\font plus
\BIBentryALTinterwordstretchfactor\fontdimen3\font minus
  \fontdimen4\font\relax}
\providecommand{\BIBforeignlanguage}[2]{{%
\expandafter\ifx\csname l@#1\endcsname\relax
\typeout{** WARNING: IEEEtran.bst: No hyphenation pattern has been}%
\typeout{** loaded for the language `#1'. Using the pattern for}%
\typeout{** the default language instead.}%
\else
\language=\csname l@#1\endcsname
\fi
#2}}
\providecommand{\BIBdecl}{\relax}
\BIBdecl

\bibitem{Scheiner2018}
N.~Scheiner, N.~Appenrodt, J.~Dickmann, and B.~Sick, ``{Radar-based Feature
  Design and Multiclass Classification for Road User Recognition},'' in
  \emph{2018 IEEE Intelligent Vehicles Symposium (IV)}.\hskip 1em plus 0.5em
  minus 0.4em\relax Changshu, China: IEEE, jun 2018, pp. 779--786.

\bibitem{Winterling2017}
T.~Winterling, J.~Lombacher, M.~Hahn, J.~Dickmann, and C.~W{\"{o}}hler,
  ``{Optimizing labelling on radar-based grid maps using active learning},'' in
  \emph{2017 18th International Radar Symposium (IRS)}, 2017, pp. 1--6.

\bibitem{Piewak2018}
F.~Piewak, P.~Pinggera, M.~Sch{\"a}fer, D.~Peter, B.~Schwarz, N.~Schneider,
  D.~Pfeiffer, M.~Enzweiler, and J.~M. Z{\"o}llner, ``Boosting lidar-based
  semantic labeling by cross-modal training data generation,'' \emph{CoRR},
  vol. abs/1804.09915, 2018.

\bibitem{richards2005}
M.~A. Richards, \emph{Fundamentals of Radar Signal Processing}, ser.
  Professional Engineering.\hskip 1em plus 0.5em minus 0.4em\relax McGraw-Hill,
  2005.

\bibitem{Agrawal2013}
L.~Agrawal and D.~Toshniwal, ``Smart phone based indoor pedestrian localization
  system,'' in \emph{2013 13th International Conference on Computational
  Science and its Applications}, June 2013, pp. 137--143.

\bibitem{Ledergerber2015}
A.~Ledergerber, M.~Hamer, and R.~D'Andrea, ``A robot self-localization system
  using one-way ultra-wideband communication,'' in \emph{2015 IEEE/RSJ
  International Conference on Intelligent Robots and Systems (IROS)}, Sept
  2015, pp. 3131--3137.

\bibitem{Thomas2010}
T.~Pany, \emph{Navigation Signal Processing for GNSS Software Receivers}.\hskip
  1em plus 0.5em minus 0.4em\relax Artech House Publishers, jan 2010.

\bibitem{Kalman1960}
R.~Kalman, ``A new approach to linear filtering and prediction problems,''
  \emph{Transactions of the ASME - Journal of Basic Engineering}, vol.~82, pp.
  35--45, jan 1960.

\end{thebibliography}
 
\end{document}